\documentclass
[superscriptaddress,secnumarabic,amssymb,amsmath,nobibnotes,aps,prd,showkeys,showpacs,nofootinbib,onecolumn,notitlepage,12pt]{revtex4}%
\usepackage{setspace}
\usepackage{color}
\usepackage{amsmath}
\usepackage{amsfonts}
\usepackage{verbatim}
\usepackage{amssymb}
\usepackage{graphicx,bm}
\usepackage{amsmath}
\usepackage{amssymb}
\usepackage{graphicx}
\usepackage{pdflscape}
\usepackage[caption=false]{subfig}%
\providecommand{\U}[1]{\protect\rule{.1in}{.1in}}
\newcommand{\be}{\begin{equation}}
\newcommand{\ee}{\end{equation}}

\newcommand{\mincir}{\raise
-3.truept\hbox{\rlap{\hbox{$\sim$}}\raise4.truept\hbox{$<$}\ }}
\newcommand{\magcir}{\raise
-3.truept\hbox{\rlap{\hbox{$\sim$}}\raise4.truept\hbox{$>$}\ }}

\ifx\pdfoutput\relax\let\pdfoutput=\undefined\fi
\newcount\msipdfoutput
\ifx\pdfoutput\undefined\else
\ifcase\pdfoutput\else
\ifx\paperwidth\undefined\else
\ifdim\paperheight=0pt\relax\else\pdfpageheight\paperheight\fi
\ifdim\paperwidth=0pt\relax\else\pdfpagewidth\paperwidth\fi
\fi\fi\fi
\begin{document}
\title{Compactification of Anisotropies in Einstein-Scalar-Gauss-Bonnet Cosmology}
\author{Alex Giacomini}
\email{alexgiacomini@uach.cl }
\affiliation{Instituto de Ciencias F\'{\i}sicas y Matem\'{a}ticas, Universidad Austral de
Chile, Valdivia, Chile}
\author{Andronikos Paliathanasis}
\email{anpaliat@phys.uoa.gr}
\affiliation{Institute of Systems Science \& Department of Mathematics, Faculty of Applied
Sciences, Durban University of Technology, Durban 4000, South Africa}
\affiliation{School for Data Science and Computational Thinking, Stellenbosch University,
44 Banghoek Rd, Stellenbosch 7600, South Africa}
\affiliation{Departamento de Matem\'{a}ticas, Universidad Cat\'{o}lica del Norte, Avda.
Angamos 0610, Casilla 1280 Antofagasta, Chile}
\affiliation{National Institute for Theoretical and Computational Sciences (NITheCS), South Africa.}
\author{Alexey Toporensky}
\email{atopor@rambler.ru}
\affiliation{Sternberg Astronomical Institute, Lomonosov Moscow State University, 119991
Moscow, Russia}

\begin{abstract}
We investigate the evolution of anisotropies in Einstein-Gauss-Bonnet theory
with a scalar field coupled to the Gauss-Bonnet term. Specifically, we examine
the simplest scenario in which the scalar field lacks a kinetic term, and its
kinetic contribution arises from an integration by parts of the Gauss-Bonnet
scalar. We consider four- and five-dimensional anisotropic spacetimes,
focusing on Bianchi I and extended Bianchi I geometries. Our study reveals
that the asymptotic solutions correspond to locally symmetric spacetimes where
at least two scale factors exhibit analogous behavior or, alternatively, to
isotropic configurations where all scale factors evolve identically.
Additionally, we discuss the effects of a cosmological constant, finding that
the presence of the cosmological constant does not lead to an isotropic universe.

\end{abstract}
\maketitle
\date{\today}

\section{Introduction}

\label{sec1}

Lovelock's theory \cite{lov1} is the generalization of General Relativity in
higher-order dimensional geometries. Lovelock theory is a second-order
gravitational theory free from Ostrogradsky instabilities and is reduced to
General Relativity in the case of a four-dimensional manifold. The
gravitational Action in Lovelock theory is constructed from geometric scalars
formed from contractions of the Riemann tensor, with the Ricci scalar being
the first construction and the Gauss-Bonnet scalar being the second
construction term.

The gravitational theory where the Lagrangian is a linear combination of the
first and second construction terms of Lovelock's theory is known as the
Gauss-Bonnet theory of gravity \cite{pan1}. Although the nonlinearity of the
gravitational field equations increases when the Gauss-Bonnet scalar appears,
the gravitational field equations remain of second order. Due to the existence
of the new geometrodynamical terms, new physical behaviors occur.
Five-dimensional black holes and wormholes within Gauss-Bonnet gravity were
investigated in \cite{tr1}; brane cosmologies with nontrivial bulk solutions
were found in \cite{ch1}. In \cite{bg1}, Birkhoff's theorem was discussed in
detail in Gauss-Bonnet and Lovelock theories. Recently, in \cite{di1}, the
phenomena of gravitational collapse in pure Gauss-Bonnet gravity were
investigated for higher-dimensional spacetimes. It was found that for a seven
dimensional geometry, the theory has gravitational dynamics indistinguishable
from Einstein's theory in the case of four dimensions, while for dimensions
five and six, gravity becomes weaker, and for dimensions higher than seven
gravity becomes stronger. For a five-dimensional geometry, the model provides
the formation of a massive timelike singularity not provided by General
Relativity \cite{di2}. Recently, a charged black hole solution was found in
\cite{di3}. On the other hand, in \cite{gbc1}, the pure Gauss-Bonnet gravity
has been used as a dark energy candidate, where the acceleration of the
universe is driven by the Gauss-Bonnet scalar. The evolution of the
anisotropies in the framework of five- and six-dimensional Bianchi I
geometries was studied in detail in \cite{top2,top1}, where new singular
behaviors of the geometry were determined.

The Gauss-Bonnet scalar is a topological invariant for a four-dimensional
geometry, which means that it does not introduce any dynamical term in the
gravitational theory. The introduction of a scalar field
\cite{gbs1,gbs2,gbs3,gbs4,gbs5} non-minimally coupled to gravity overcomes
this property, and this new gravitational theory remains of second order while
the new degrees of freedom lead to new behavior for the physical variables and
to new phenomena, like new inflationary behavior, and new strong gravitational
field solutions; for more details, we refer the reader to
\cite{gg1,gg2,gg2a,gg3,gg4,gg5,gg6,gg7,gg8,gg9,gg10,gg11,gg12,g11,g12,g13,g14} and references
therein. Another attempt is the introduction of a nonlinear function of the
Gauss-Bonnet term in the modification of the Einstein-Hilbert Action
\cite{gbm01,gbm03}. Nevertheless, by introducing a Lagrange multiplier, the
latter theory can have a scalar field description as in \cite{gbs1}. In
\cite{agb1} it was found that there exist a spontaneous symmetry breaking as a
result of extra symmetry breaking within the Einstein-Gauss-Bonnet scalar
field theory when the scalar field is coupled not to the Gauss-Bonnet term but
to the Ricci scalar.

In this study, we investigate the evolution of the anisotropies within the
Einstein-Gauss-Bonnet scalar field theory. In particular, we consider a
four-dimensional Bianchi I geometry and we perform a detailed study for the
phase-space by introducing dimensionless variables. We focus on the case where
there is not any kinetic term for the scalar field in the gravitational
Lagrangian. This is a very interesting model because it shows that there exist
anisotropic solutions which are decomposable, that is, at least one of the
scale factors is constant, and the isometries of the Bianchi geometry can
become gradient isometries. At the same time, the asymptotic solutions admit
more than three isometries. The isotropic universe is an attractor for the
model, but for general initial conditions, it is possible to have attractors
which describe Big Rip singularities. The introduction of the cosmological
constant dramatically changes the behavior of the asymptotic solutions, but
now we can have future solutions which lead to anisotropic decomposable
spacetimes. Finally, in order to understand the behavior of the anisotropies
in higher-dimensional spacetimes, we present some numerical solutions in the
case of a five-dimensional Bianchi I (-like) geometry. We see that new
behavior exists, while there are initial conditions in which the pure
Gauss-Bonnet solution can be recovered. The structure of the paper is as
described below.

In Section \ref{sec2}, we examine the gravitational model under consideration,
which is Einstein-Scalar-Gauss-Bonnet gravity in a four-dimensional manifold.
Here, the scalar field is coupled to the Gauss-Bonnet scalar to ensure a
nonzero contribution to the gravitational field equations. We use the Bianchi
I line element with three distinct scale factors and derive the point-like
Lagrangian for the field equations as provided by the minisuperspace approach.
Sections \ref{sec3} and \ref{sec4d} contains the main results of this study,
where we conduct a detailed analysis of the asymptotic behavior of the
physical variables.

Moreover, in Section \ref{sec3}, we assume that the scalar field appears in
the gravitational model solely through its coupling with the Gauss-Bonnet
scalar. Given that the Gauss-Bonnet scalar is a boundary term, integration by
parts introduces kinetic components for the scalar field in the field
equations, resulting in second-order derivatives of the scalar field. For this
model, the asymptotic analysis reveals a unique stationary point describing an
isotropic universe, while the other stationary points lead to anisotropic
universes with two equal scale factors. There are also points with one or two
constant scale factors. Introducing the cosmological constant yields similar
outcomes; however, there is no isotropic solution with a nonzero cosmological
constant. In Section \ref{sec4d}, we extend our consideration to the case of a
five-dimensional anisotropic background geometry. Our findings and conclusions
are summarized in Section \ref{sec5}.

\section{Anisotropic Einstein-Scalar-Gauss-Bonnet Gravity}

\label{sec2}

We consider the Einstein-Gauss-Bonnet Action, with a scalar-field coupled to
the Gauss-Bonnet scalar, that is,%
\begin{equation}
S=\int d^{4}x\sqrt{-g}\left(  R-\phi G-2\Lambda\right)  , \label{ai.01}%
\end{equation}
where $R$ is the Ricci scalar of the four-dimensional Riemannian space with
metric tensor $g_{\mu\nu}$, $G$ is the Gauss-Bonnet scalar and $\phi$ is the
scalar field coupling to the Gauss-Bonnet term, parameter $\Lambda$ is the
cosmological constant. \ 

The Gauss-Bonnet scalar is defined as
\begin{equation}
G=R^{2}-4R_{\mu\nu}R^{\mu\nu}+R_{\mu\nu\kappa\lambda}R^{\mu\nu\kappa\lambda}.
\label{ai.02}%
\end{equation}
where $R_{\mu\nu}$ is the Ricci tensor and~$R_{\mu\nu\kappa\lambda}$ the
Riemann tensor for the metric $g_{\mu\nu}$. Nevertheless, for a
four-dimensional space, scalar $G$ is a pure boundary term, when $\phi$ is a
nonconstant function, then the component $\phi G$ of the gravitational
Lagrangian contributes to the cosmic evolution. We remark that without loss of
generality we can assume the scalar field $\varphi$ with coupling function
$\phi=\phi\left(  \varphi\right)  $, thus, since there is not kinetic term or
potential function the analysis is independent from the nonlinear definition
of function $\phi\left(  \varphi\right)  $.

One might assume that, in this context, the scalar field is non-dynamical.
However, this is not the case. As we will show below, a kinetic component for
the scalar field emerges from the variation with respect to the metric tensor
due to the boundary contributions of the Gauss-Bonnet scalar.

\subsection{Bianchi I background}

We introduce the anisotropic Bianchi I spacetime with line element%
\begin{equation}
ds^{2}=-N^{2}\left(  t\right)  dt^{2}+\left(  S_{1}\left(  t\right)  \right)
^{2}dx^{2}+\left(  S_{2}\left(  t\right)  \right)  ^{2}dy^{2}+\left(
S_{3}\left(  t\right)  \right)  ^{2}dz^{2}, \label{ai.03}%
\end{equation}
where $N\left(  t\right)  $ is the lapse function and $S_{1}\left(  t\right)
,~S_{2}\left(  t\right)  $ and $S_{3}\left(  t\right)  $ are the three scale
factors. The volume of the three-dimensional hypersurface is defined as
$V=S_{1}\left(  t\right)  S_{2}\left(  t\right)  S_{3}\left(  t\right)  $.
From the three scale factors we can define the three Hubble functions%
\begin{equation}
H_{1}=\frac{\dot{S}_{1}}{S},~H_{2}=\frac{\dot{S}_{2}}{S_{2}},~H_{3}=\frac
{\dot{S}_{3}}{S_{3}}. \label{ai.03a}%
\end{equation}

In terms of the Misner variables the line element (\ref{ai.03}) reads%
\begin{equation}
ds^{2}=-N^{2}\left(  t\right)  dt^{2}+e^{3a}\left(  e^{2\beta_{+}\left(
t\right)  }dx^{2}+e^{-\beta_{+}\left(  t\right)  }\left(  e^{\beta_{-}\left(
t\right)  }dy^{2}+e^{-\beta_{-}\left(  t\right)  }dz^{2}\right)  \right)  ,
\label{ai.04}%
\end{equation}
where now the volume is defined as $V\left(  t\right)  =e^{3a}$ and $\beta
_{+}\left(  t\right)  $,~$\beta_{-}\left(  t\right)  $ are the two anisotropic
parameters. When $\beta_{+}\left(  t\right)  =\beta_{\_}\left(  t\right)  =0$,
the spacetime is reduced to the isotropic spatially flat FLRW geometry. The
Hubble function is defined as $H=\frac{\dot{a}}{a}$, such that
\begin{equation}
H=\frac{1}{3}\left(  H_{1}+H_{2}+H_{3}\right)  , \label{ai.04a}%
\end{equation}
or%
\begin{align}
H_{1}  &  =H\left(  1+\Sigma_{1}\right)  ,~\\
H_{2}  &  =H\left(  1-\frac{1}{2}\left(  \Sigma_{+}-\sqrt{3}\Sigma_{2}\right)
\right)  ,\\
H_{3}  &  =H\left(  1-\frac{1}{2}\left(  \Sigma_{+}+\sqrt{3}\Sigma_{2}\right)
\right)  .
\end{align}
are the expansion rates on the three Killing directions.

From the line-element (\ref{ai.04}) we calculate the Ricciscalar
\begin{equation}
R=6\ddot{a}+12\dot{a}^{2}+\frac{3}{2}\left(  \dot{\beta}_{+}\right)
^{2}+\frac{1}{2}\left(  \dot{\beta}_{-}\right)  ^{2}, \label{ai.05}%
\end{equation}
and%
\begin{equation}
\int Ne^{3a}Gdt=\frac{2}{N^{3}}e^{3a}\left(  \dot{a}+\dot{\beta}_{+}\right)
\left(  2\dot{a}-\dot{\beta}_{+}-\dot{\beta}_{-}\right)  \left(  2\dot{a}%
-\dot{\beta}_{-}+\dot{\beta}_{+}\right)  . \label{ai.06}%
\end{equation}

\subsection{Minisuperspace description}

By replacing (\ref{ai.05}) and (\ref{ai.06}) in the Action Integral
(\ref{ai.01}) and integrating by parts we end with the following point-like
Lagrangian function%
\begin{align}
L\left(  N,a,\dot{a},\beta_{\pm},\dot{\beta}_{\pm}\right)   &  =\frac{e^{3a}%
}{N}\left(  -3\dot{a}^{2}+\frac{3}{4}\dot{\beta}_{+}^{2}+\frac{1}{4}\dot
{\beta}_{-}^{2}-2\Lambda N\right) \nonumber\\
&  ~~~~~+\left(  \frac{e^{3\alpha}}{N^{3}}\dot{\phi}\left(  \dot{a}+\dot
{\beta}_{+}\right)  \left(  2\dot{a}-\dot{\beta}_{+}-\dot{\beta}_{-}\right)
\left(  2\dot{a}-\dot{\beta}_{-}+\dot{\beta}_{+}\right)  \right)  .
\label{ai.07}%
\end{align}
We remark that when $\dot{\phi}=0$, i.e. $\phi=0$, then the limit of General
Relativity is recovered.

The field equations follow from the variation of the aforementioned~point-like
Lagrangian function with respect to the dynamical variables $N,~a$,
$\beta_{\pm}$ and $\phi\,$.

Without loss of generality we assume the lapse function $N=1$. Thus, for the
constant lapse function the gravitational field equations read%
\begin{equation}
0=3H^{2}-\frac{3}{4}\dot{\beta}_{+}^{2}-\frac{1}{4}\dot{\beta}_{-}^{2}%
-3\dot{\phi}\left(  H+\dot{\beta}_{+}\right)  \left(  2H-\dot{\beta}_{+}%
-\dot{\beta}_{-}\right)  \left(  2H-\dot{\beta}_{-}+\dot{\beta}_{+}\right)
-2\Lambda, \label{ai.08}%
\end{equation}

\begin{align}
0  &  =2\dot{H}+H^{2}\left(  3-4\ddot{\phi}\right)  -2\Lambda+\dot{\beta}%
_{+}^{3}\dot{\phi}-8H\dot{H}\dot{\phi}-\dot{\beta}_{+}\dot{\phi}\left(
\dot{\beta}_{-}^{2}-2\ddot{\beta}_{+}\right) \nonumber\\
&  ~~~+\dot{\beta}_{+}^{2}\left(  \frac{3}{4}+\ddot{\phi}\right)  +\frac
{1}{12}\dot{\beta}_{-}\left(  \dot{\beta}_{-}\left(  3+\ddot{\phi}\right)
+8\dot{\phi}\ddot{\beta}_{-}\right)  , \label{ai.09}%
\end{align}

\begin{align}
0  &  =-\frac{3}{2}\ddot{\beta}_{+}+18H^{2}\dot{\beta}_{+}\dot{\phi}%
+2\dot{\phi}\left(  3\dot{\beta}_{+}\left(  \dot{H}-\ddot{\beta}_{-}\right)
+\dot{\beta}_{-}\ddot{\beta}_{-}\right)  +\ddot{\phi}\left(  \dot{\beta}%
_{-}^{2}-3\dot{\beta}_{+}^{2}\right) \nonumber\\
&  ~~~~~~~+\frac{3}{2}H\left(  2\dot{\phi}\left(  \dot{\beta}_{-}+2\ddot
{\beta}_{+}\right)  +\dot{\beta}_{+}\left(  4\ddot{\phi}-3\right)
-6\dot{\beta}_{+}^{2}\dot{\phi}\right)  , \label{ai.10}%
\end{align}%
\begin{align}
0  &  =\ddot{\beta}_{-}\left(  4\dot{\beta}_{+}\dot{\phi}-1\right)
+12H^{2}\dot{\beta}_{-}\dot{\phi}+4\dot{\beta}_{-}^{2}\left(  \dot{\phi
}\left(  \dot{H}+\ddot{\beta}_{+}\right)  +\dot{\beta}_{+}\ddot{\phi}\right)
\nonumber\\
&  ~~~+H\left(  4\dot{\phi}\ddot{\beta}_{-}+\dot{\beta}_{-}\left(  4\ddot
{\phi}-3+12\dot{\beta}\dot{\phi}\right)  \right)  , \label{ai.11}%
\end{align}
and%
\begin{equation}
0=\left(  e^{3\alpha}\left(  H+\dot{\beta}_{+}\right)  \left(  2H-\dot{\beta
}_{+}-\dot{\beta}_{-}\right)  \left(  2H-\dot{\beta}_{-}+\dot{\beta}%
_{+}\right)  \right)  ^{\cdot} \label{ai.12}%
\end{equation}

In the following Section we study the dynamical evolution of the anisotropic
parameters and the evolution of the physical parameters. We employ
dimensionless variables and we perform a detailed phase-space analysis.

\section{Evolution of the anisotropies}

\label{sec3}

We work within the framework of the $H$-normalization \cite{cop1} such we
introduce the new dependent\ dimensionless variables
\begin{equation}
x=H\dot{\phi},~\Sigma_{+}=\frac{\dot{\beta}_{+}}{H},~\Sigma_{-}=\frac
{\dot{\beta}_{-}}{\sqrt{3}H}~,~\Omega_{\Lambda}=\frac{\Lambda}{3H^{2}},
\label{ai.14}%
\end{equation}
and the independent variable $\tau=\ln a$. At this point we remark that from
(\ref{ai.09}) it is possible the Hubble function $H$ to change sign, that is,
to take the value $H=0$; in the theories with this properties another
normalization approach is considered (see for instance \cite{gg9,gg10});
however in this model the consideration of the more general normalization does
not lead to the detection of new solutions with different physical properties.
Hence, without loss of physical properties, we focus with the $H$%
-normalization and we assume that $H>0$ (or $H<0$).

With the use of the dimensionless variables, the gravitational field equations
reduced to an algebraic-differential system of the form%
\begin{equation}
\frac{d}{d\tau}\mathbf{A=F}\left(  \mathbf{A}\right)  \label{ai.15}%
\end{equation}
where $\mathbf{A}=\left(  x,\Sigma_{+},\Sigma_{-},\Omega_{\Lambda}\right)  $
and $\mathbf{F}=\left(  F_{1},F_{2},F_{3},F_{4}\right)  $. The algebraic
constraint equation reads%
\begin{equation}
4\left(  1-2\Omega_{\Lambda}\right)  -\Sigma_{+}^{2}-\Sigma_{-}^{2}-4x\left(
1+\Sigma_{+}\right)  \left(  \left(  2-\Sigma_{+}\right)  ^{2}-3\Sigma_{-}%
^{2}\right)  =0. \label{ai.16}%
\end{equation}

\subsection{Model $\Lambda=0$}

In the absence of the cosmological constant, i.e. on the surface where
$\Omega_{\Lambda}=0$, the dynamical system (\ref{ai.15}) reads%
\begin{align}
\frac{dx}{d\tau}  &  =\frac{1}{4}\left(  1-2x\left(  2+\Sigma_{+}^{2}%
+\Sigma_{-}^{2}\right)  \right)  ,\label{ai.17}\\
\frac{d\Sigma_{+}}{d\tau}  &  =\frac{1}{2}\left(  1+\Sigma_{+}\right)  \left(
\Sigma_{-}^{2}+\Sigma_{+}\left(  \Sigma_{+}-2\right)  \right)  ,
\label{ai.18}\\
\frac{d\Sigma_{-}}{d\tau}  &  =\frac{1}{2}\Sigma_{-}\left(  \Sigma_{-}%
^{2}+\Sigma_{+}\left(  \Sigma_{+}+2\right)  -2\right)  , \label{ai.19}%
\end{align}
while the constraint equation is simplified as
\begin{equation}
4-\Sigma_{+}^{2}-\Sigma_{-}^{2}-4x\left(  1+\Sigma_{+}\right)  \left(  \left(
2-\Sigma_{+}\right)  ^{2}-3\Sigma_{-}^{2}\right)  =0. \label{ai.20}%
\end{equation}
With the use of the the algebraic equation (\ref{ai.20}) the dimension of the
dynamical system is reduced to two. We observe that the dynamical system is
foliated, and there is not any dependence of the dynamical system of the
anisotropic variables $\Sigma_{+}$,~$\Sigma_{-}$ from the value of variable
$x$. The only dependency follows from the constraint (\ref{ai.20}).

Each point $P=\left(  x\left(  P\right)  ,\Sigma_{+}\left(  P\right)
,\Sigma_{-}\left(  P\right)  \right)  $, describes a universe where the
deceleration parameter $q=-1-\frac{\dot{H}}{H^{2}}$ is given by the expression%
\begin{equation}
q\left(  P\right)  =\frac{1}{2}\left(  \Sigma_{+}^{2}\left(  P\right)
+\Sigma_{-}^{2}\left(  P\right)  \right)  .
\end{equation}
Thus, $q\left(  P\right)  \geq0$, where $q\left(  P\right)  =0$ only in the
isotropic limit of FLRW.

In order to understand the asymptotic behaviour of the dynamical system
(\ref{ai.17})-(\ref{ai.20}) we determine the stationary points. For each
stationary point we recover the physical solution and we investigate the
stability properties.

The stationary points for the dynamical system (\ref{ai.17})-(\ref{ai.20}) are%
\[
P_{1}=\left(  \frac{1}{4},0,0\right)  ,
\]
from where we derive $q\left(  P_{1}\right)  =0$. The solution describes an
isotropic FLRW geometry with Hubble function $H\left(  t\right)  =\frac{1}{t}$
and volume $e^{3a\left(  t\right)  }=\left(  a_{0}t\right)  ^{3}$. The unique
scale factor is $e^{a\left(  t\right)  }=a_{0}t$. In order to determine the
stability properties for the stationary point we determine the eigenvalues of
the linearized system around the stationary point for the two equations
(\ref{ai.18}), (\ref{ai.19}). We calculate the negative eigenvalues $-1$,~$-1$
\ from where we infer that the point is an attractor.
\[
P_{2}=\left(  \frac{1}{12},2,0\right)  ,~P_{3}=\left(  \frac{1}{12}%
,-1,\sqrt{3}\right)  ,~P_{4}=\left(  \frac{1}{12},-1,-\sqrt{3}\right)  .
\]
with $q\left(  P_{2,3,4}\right)  =2$, that is $H\left(  t\right)  =\frac
{1}{3t}~$and volume $e^{3a\left(  t\right)  }=\left(  a_{0}\right)  ^{3}t$.
The anisotropic parameters are $\left(  \beta_{+},\beta_{-}\right)  _{P_{2}%
}=\left(  \frac{2}{3t},0\right)  $, $\left(  \beta_{+},\beta_{-}\right)
_{P_{3}}=\left(  -\frac{1}{3t},\frac{3}{t}\right)  $ and $\left(  \beta
_{+},\beta_{-}\right)  _{P_{4}}=\left(  -\frac{1}{3t},-\frac{3}{t}\right)  $.~\ 

Therefore, the each of the above points the three Hubble functions for the
line element (\ref{ai.03}) are calculated $\left(  H_{1},H_{2},H_{3}\right)
_{P_{2}}=\left(  \frac{1}{t},0,0\right)  $;~$\left(  H_{1},H_{2},H_{3}\right)
_{P_{3}}=\left(  0,\frac{1}{t},0\right)  $ and $\left(  H_{1},H_{2}%
,H_{3}\right)  _{P_{4}}=\left(  0,0,\frac{1}{t}\right)  $. Thus the
corresponding scale factors are derived $\left(  S_{1},S_{2},S_{3}\right)
_{P_{2}}=\left(  S_{10}t,0,0\right)  $; $\left(  S_{1},S_{2},S_{3}\right)
_{P_{2}}=\left(  0,S_{20}t,0\right)  $ and~$\left(  S_{1},S_{2},S_{3}\right)
_{P4}=\left(  0,0,S_{30}t\right)  $. For these three stationary points we
derive the positive eigenvalues $+3,~+3$, thus points $P_{2}$,~$P_{3}$ and
$P_{4}$ are characterized as sources and they describe unstable
solutions.\ \ These three points describe Kanser spacetimes where the two of
the Kasner indices are zero, and the third is one, that is, the spacetime is
the flat space.

Indeed, consider the asymtotpic solution described by point $P_{2}$, then, the
line element reads%
\begin{equation}
ds^{2}=-dt^{2}+t^{2}dx^{2}+dy^{2}+dz^{2},
\end{equation}
where without loss of generalizy we assumed that $S_{10}=1,~S_{20}=1$ and
$S_{30}=1$. After the change of variables $T=t\cosh^{2}x$,$~X=t\sinh^{2}%
x\,\ $the line element takes the diagonal form%
\begin{equation}
ds^{2}=-dT^{2}+dX^{2}+dy^{2}+dz^{2}.
\end{equation}

\subsubsection{Compactified variables}

The dynamical variables $x,~\Sigma_{\pm}$ are not constraint, which means that
they can take values at the infinity.

In order to study the dynamical evolution at the infinity we define the
compactified variables%
\begin{equation}
\left(  \Sigma_{+},\Sigma_{-}\right)  =\left(  \frac{Y_{+}}{\sqrt{1-Y_{+}%
^{2}-Y_{-}^{2}}},\frac{Y_{-}}{\sqrt{1-Y_{+}^{2}-Y_{-}^{2}}}\right)
,~dT=\sqrt{1-Y_{+}^{2}-Y_{-}^{2}}d\tau,
\end{equation}
therefore the two-dimensional dynamical system (\ref{ai.18}), (\ref{ai.19})
reads%
\begin{align}
\frac{dY_{+}}{dT}  &  =\frac{1}{2}\left(  Y_{+}^{4}+Y_{-}^{2}-Y_{+}^{2}\left(
1+3Y_{-}^{2}\right)  +Y_{+}\left(  3\left(  Y_{+}^{2}+Y_{-}^{2}\right)
-2\right)  \sqrt{1-Y_{+}^{2}-Y_{-}^{2}}\right)  ,\label{ds.01}\\
\frac{dY_{-}}{dT}  &  =\frac{1}{2}Y_{-}\left(  Y_{+}^{3}+Y_{+}\left(
2-3Y_{-}^{2}\right)  +\left(  3\left(  Y_{+}^{2}+Y_{-}^{2}\right)  -2\right)
\sqrt{1-Y_{+}^{2}-Y_{-}^{2}}\right)  . \label{ds.02}%
\end{align}

The stationary points $Q=\left(  Y_{+}\left(  Q\right)  ,Y_{-}\left(
Q\right)  \right)  ~$for the latter dynamical system are~%
\[
Q_{1}=\left(  0,0\right)  ,~Q_{2}=\left(  \frac{2}{\sqrt{5}},0\right)
,~Q_{3}^{\pm}=\frac{1}{\sqrt{5}}\left(  -1,\pm\sqrt{3}\right)  ,~Q_{4}=\left(
-\frac{1}{\sqrt{2}},0\right)  ,~
\]%
\[
Q_{5}^{\pm}=\frac{1}{2\sqrt{2}}\left(  1,\pm\sqrt{3}\right)  ,~Q_{6}^{\pm
}=\left(  \pm1,0\right)  ,~Q_{7}^{\pm}=\frac{1}{2}\left(  1,\pm\sqrt
{3}\right)  ~,~Q_{8}^{\pm}=\frac{1}{2}\left(  -1,\pm\sqrt{3}\right)  .
\]
Stationary points $Q_{1}$,~$Q_{2}$ and $Q_{3}^{\pm}$ are points $P_{1}%
$,~$P_{2}$, $P_{3}$ and $P_{4}$ respectively, defined in the finite regime.

Stationary points \ $Q_{4}$ and $Q_{5}^{\pm}$ are points in which the variable
$x$ reaches infinity. The deceleration parameter is derived $q\left(
Q_{4}\right)  =\frac{1}{2}$ and $q\left(  Q_{5}^{\pm}\right)  =\frac{1}{2}$,
that is, $H\left(  t\right)  =\frac{2}{3t}$ and $e^{a\left(  t\right)  }%
=a_{0}t^{\frac{2}{3}}$. Furthermore, the anisotropic parameters $\Sigma_{\pm}$
are $\left(  \Sigma_{+},\Sigma_{-}\right)  _{Q_{4}}=\left(  -1,0\right)  $;
$\left(  \Sigma_{+},\Sigma_{-}\right)  _{Q_{5}^{\pm}}=\left(  \frac{1}{2}%
,\pm\frac{\sqrt{3}}{2}\right)  $. Hence, for the line element (\ref{ai.03}) we
determine the Hubble functions $\left(  H_{1},H_{2},H_{3}\right)  _{Q_{4}%
}=\left(  0,\frac{1}{t},\frac{1}{t}\right)  $; $\left(  H_{1},H_{2}%
,H_{3}\right)  _{Q_{5}^{+}}=\left(  \frac{1}{t},\frac{1}{t},0\right)  $ and
$\left(  H_{1},H_{2},H_{3}\right)  _{Q_{5}^{-}}=\left(  \frac{1}{t},0,\frac
{1}{t}\right)  $. The corresponding scale factors are $\left(  S_{1}%
,S_{2},S_{3}\right)  _{Q_{4}}=\left(  0,t,t\right)  $ ; $\left(  S_{1}%
,S_{2},S_{3}\right)  _{Q_{5}^{+}}=\left(  t,t,0\right)  $ and $\left(
S_{1},S_{2},S_{3}\right)  _{Q_{5}^{-}}=\left(  t,0,t\right)  $. These
stationary points describe LRS Bianchi I geometries with one scale factor constant.

Finally, the rest of the stationary points are on the surface $1-Y_{+}%
^{2}-Y_{-}^{2}=0$, where the anisotropic parameters take values at infinity.
The deceleration parameter reads $q=\frac{1}{2}\left(  -1+\frac{1}{1-Y_{+}%
^{2}-Y_{-}^{2}}\right)  ;$ hence for these stationary points $q\rightarrow
+\infty$ which means that these points describe Big Crunch singularities,
where two of the Hubble functions are equal.

As far as the stability properties of the new points is concerned, point
$Q_{4}$ leads to the eigenvalues $-\frac{3}{2\sqrt{2}},~+\frac{3}{2\sqrt{2}}$,
thus $Q_{4}$ is a saddle point. For points $Q_{5}^{\pm}$ we calculate
$-\frac{3}{4\sqrt{2}}\sqrt{1\mp\sqrt{3}},~-\frac{3}{4\sqrt{2}}\sqrt{1\mp
\sqrt{3}}$, that is, $\ Q_{5}^{+}$ is a center point and $Q_{5}^{-}$ is a
saddle point. As far as the rest of the points at the infiniy, it follows that
$Q_{6}^{-}$, $Q_{7}^{+}$ and $Q_{8}^{-}$ are attractors, while the rest of the
points describe unstable solutions.

The equilibrium points and their physical interpretation are summarized in
Table \ref{tab1}.

\begin{figure}[ptbh]
\centering\includegraphics[width=0.8\textwidth]{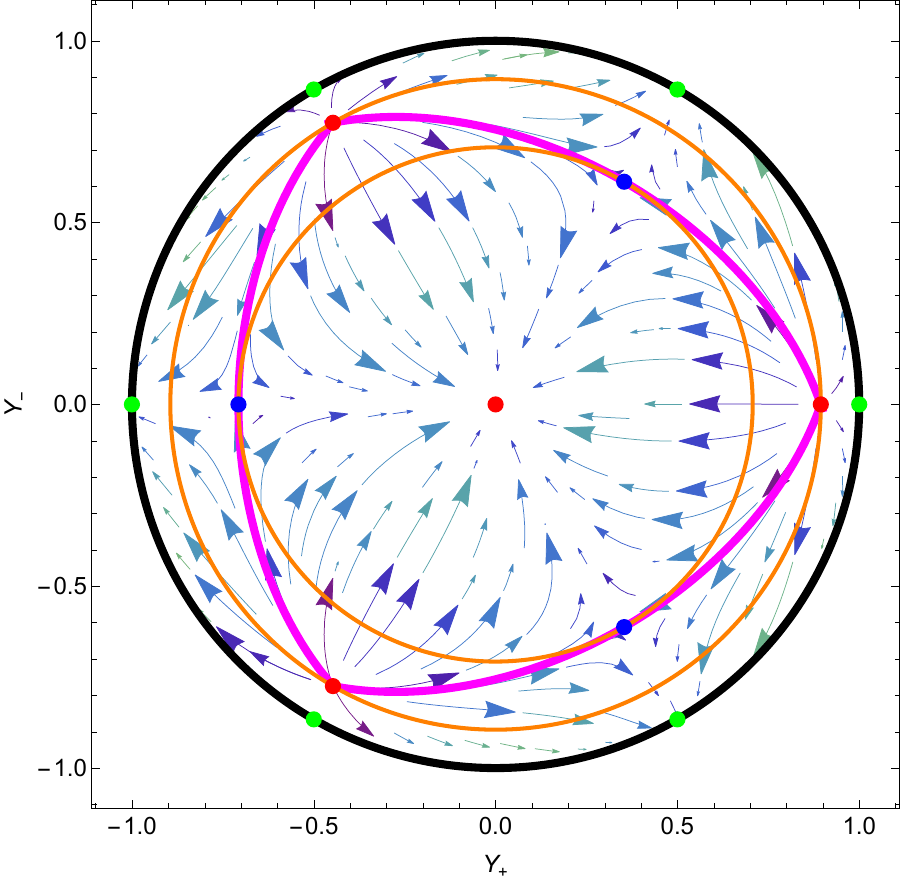}\caption{Phase-space
portrait for the dynamical system (\ref{ds.01}), (\ref{ds.02}) in the
compactified variables $Y_{+}$ and $Y_{-}$. With red are marked the points
$P_{1}$,~$P_{2}$,~$P_{3}$ and $P_{4}$, with blue the points $Q_{4}$ and
$Q_{5}^{\pm}$. With green are marked the rest of the stationary points at the
infinity. The black circle denotes the infinity line, while the two orange
circles describe define the geometric space of points $P_{1}$, $P_{2}$,
$P_{3}$, $P_{4}$ and $Q_{4},~Q_{5}^{\pm}$. We observe that for initial
conditions inside the area defined by the geometric space (magenta) which
connect the stationary points $P_{2}$, $P_{3}$, $P_{4}$ and $Q_{4},~Q_{5}%
^{\pm}$ the future attractor is always the isotropic FLRW universe. This
geometric space is defined by three ellipses.}%
\label{ff1}%
\end{figure}

In Fig. \ref{ff1} we plot the phase-space portrait for the dynamical system
(\ref{ds.01}), (\ref{ds.02}). We observe that for initial conditions inside
the area defined by the geometric space which connect the stationary points
$P_{2}$, $P_{3}$, $P_{4}$ and $Q_{4},~Q_{5}^{\pm}$ the future attractor is
always the isotropic FLRW universe; otherwise the future attractor is the
anisotropic LRS Bianchi universe described by the points at the infinity. This
geometric space is the interception of the ellipse with equation%
\begin{equation}
2x^{2}+y^{2}=1,
\end{equation}
and the two other ellipses which are the rotations by $\frac{\pi}{3}$ and
$\frac{2\pi}{3}$ of the latter ellipse.

The latter stationary points at the infinity where $q\rightarrow+\infty,$
describe gravitational collapse; however, by definition we have assumed that
$H$ can not change sign. Hence, if $H>0$, these points indicate that
asymptotically we reach asymptotic solutions where $H$ reaches zero, such $H$
to change sing. From Fig. \ref{ff1} it is clear that in order to avoid such
behaviour the initial conditions of the problem should be defined within the
space defined by the latter ellipses.%

%TCIMACRO{\TeXButton{B}{\begin{table}[tbp] \centering}}%
%BeginExpansion
\begin{table}[tbp] \centering
%EndExpansion
\caption{Stationary points and their physical properties for the dynamical system (\ref{ai.17})-(\ref{ai.20}).}%
\begin{tabular}
[c]{ccccc}\hline\hline
\textbf{Point} & \textbf{Spacetime} & $\mathbf{q}$ & $\left(  \mathbf{S}%
_{1}\mathbf{,S}_{2}\mathbf{,S}_{3}\right)  $ & Stability\\\hline
$P_{1}$ & FLRW & $0$ & $\left(  S_{10}\,t,S_{20}t,S_{30}t\right)  $ &
Attractor\\
$P_{2}$ & LRS\ Bianchi\ I (Flat space) & $2$ & $\left(  S_{10}t,S_{20}%
,S_{30}\right)  $ & Source\\
$P_{3}$ & LRS\ Bianchi\ I (Flat space) & $2$ & $\left(  S_{10},S_{20}%
t,S_{30}\right)  $ & Source\\
$P_{4}$ & LRS\ Bianchi\ I (Flat space) & $2$ & $\left(  S_{10},S_{20}%
,S_{30}t\right)  $ & Source\\
$Q_{4}$ & LRS\ Bianchi\ I & $\frac{1}{2}$ & $\left(  S_{10},S_{20}%
t,S_{30}t\right)  $ & Saddle\\
$Q_{5}^{\pm}$ & LRS\ Bianchi\ I & $\frac{1}{2}$ & $%
\begin{array}
[c]{c}%
\left(  S_{10}t,S_{20}t,S_{30}\right)  ;~\text{or}\\
\left(  S_{10}t,S_{20},S_{30}t\right)
\end{array}
$ & Saddle\\
$Q_{6}^{\pm}$ & LRS\ Bianchi\ I & $\infty$ & Undefined & $Q_{6}^{-}$
Attractor\\
$Q_{7}^{\pm}$ & LRS\ Bianchi\ I & $\infty$ & Undefined & $Q_{7}^{+}$
Attractor\\
$Q_{8}^{\pm}$ & LRS\ Bianchi\ I & $\infty$ & Undefined & $Q_{8}^{-}$
Attractor\\\hline\hline
\end{tabular}
\label{tab1}%
%TCIMACRO{\TeXButton{E}{\end{table}}}%
%BeginExpansion
\end{table}%
%EndExpansion

\subsection{Model $\Lambda\neq0$}

Consider now the case where the cosmological constant $\Lambda$ has a nonzero
value. Before we proceed with the investigation of the stationary points we
focus with the special cases, if the isotropic solution exist; and if the
limit of General Relativity is recovered.

On the surface where $\Sigma_{+}=0$, $\Sigma_{-}=0$, that is, the spacetime is
isotropic and described the FLRW line element, the dynamical system
(\ref{ai.15}), (\ref{ai.16}) is reduced to the following system%
\begin{align}
\frac{dx}{d\tau}  &  =\frac{1}{4}\left(  1-6\Omega_{\Lambda}-4x\right)  ,\\
\frac{d\Sigma_{\pm}}{d\tau}  &  =0,
\end{align}
with algebraic constraint $3-6\Omega_{\Lambda}-12x=0$. Therefore, on this
surface it follows $x=\frac{1}{4}+x_{0}e^{2t}$. Therefore there is not any
isotropic stationary point for $\Omega_{\Lambda}\neq0$. We proceed with the
investigation of the stationary points $R=\left(  x\left(  R\right)
,\Sigma_{+}\left(  R\right)  ,\Sigma_{-}\left(  R\right)  ,\Omega_{\Lambda
}\left(  R\right)  \right)  $ for $\Omega_{\Lambda}\neq0$ for the dynamical
system (\ref{ai.15}), (\ref{ai.16}) at the finite regime.

The stationary points are%
\[
R_{1}=\left(  \frac{1}{6},-1,0,\frac{3}{8}\right)  ,~R_{2}^{\pm}=\left(
\frac{1}{6},\frac{1}{2},\frac{\sqrt{3}}{2},\frac{3}{8}\right)  .
\]
For the stationary points we calculate $q\left(  R_{1}\right)  =-1$ and
$q\left(  R_{2}^{\pm}\right)  =-1$, from where it follows that $H=H_{0}$. The
three Hubble function for the line element (\ref{ai.03}) are calculated
$\left(  H_{1},H_{2},H_{3}\right)  _{R_{1}}=\frac{3}{2}\left(  0,1,1\right)
H$, $\left(  H_{1},H_{2},H_{3}\right)  _{R_{2}^{+}}=\frac{3}{2}\left(
1,1,0\right)  H$ and $\left(  H_{1},H_{2},H_{3}\right)  _{R_{2}^{-}}=\frac
{3}{2}\left(  1,0,1\right)  H$. Hence, the stationary points describe LRS
Bianchi I spacetimes, with the two scale factors to be exponential and the
third one to be constant, that is $\left(  S_{1},S_{2},S_{3}\right)  _{R_{1}%
}=\left(  S_{10},S_{20}e^{\frac{3}{2}H},S_{30}e^{\frac{3}{2}H}\right)  $,
$\left(  S_{1},S_{2},S_{3}\right)  _{R_{1}}=\left(  S_{10}e^{\frac{3}{2}%
H},S_{20}e^{\frac{3}{2}H},S_{30}\right)  $ and $\left(  S_{1},S_{2}%
,S_{3}\right)  _{R_{1}}=\left(  S_{10}e^{\frac{3}{2}H},S_{20},S_{30}%
e^{\frac{3}{2}H}\right)  $.

The linearized three-dimensional dynamical system (\ref{ai.15}), (\ref{ai.16})
around the above stationary points provide the three negative eigenvalues
$-3,$~$-3,~-3$, from where we infer that the stationary points are always
attractors and the asymptotic solutions are stable.

\begin{figure}[ptbh]
\centering\includegraphics[width=1\textwidth]{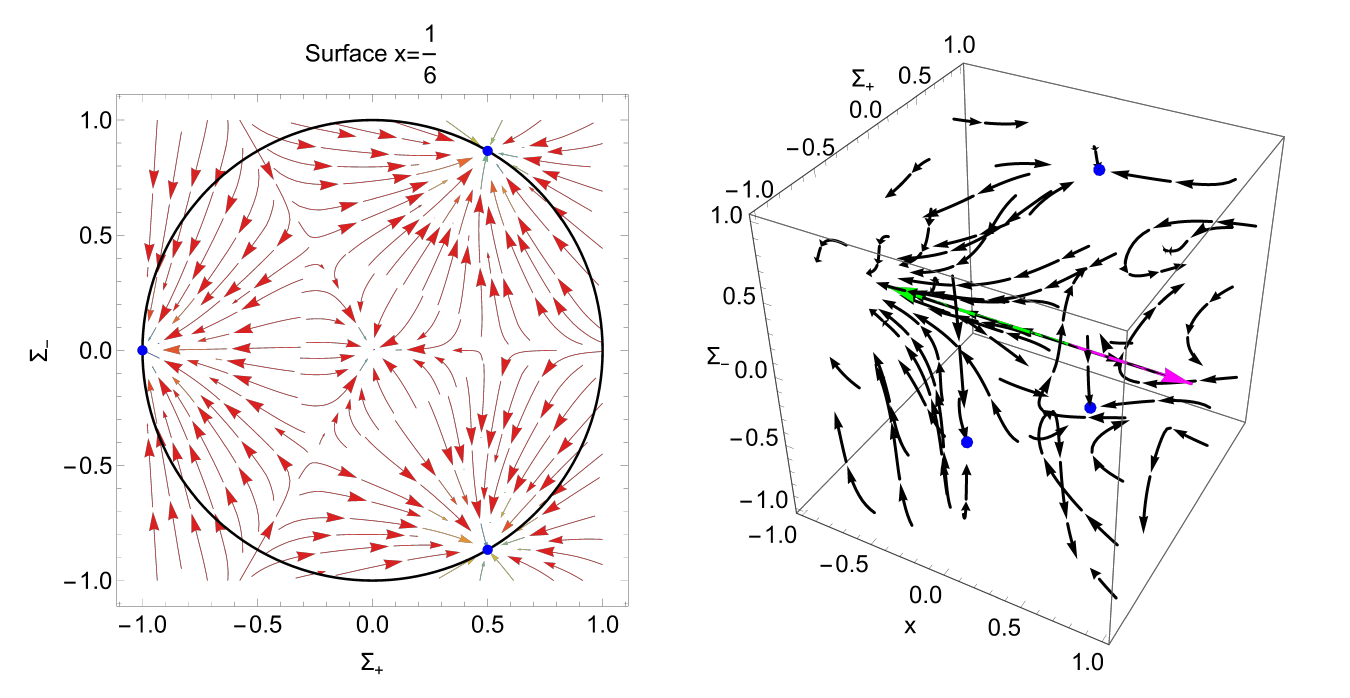}\caption{Phase-space
portrait for the three-dimensional dynamical system (\ref{ai.15}),
(\ref{ai.16}) on the space $\left(  x,\Sigma_{\pm}\right)  $. Left Fig. is the
two-dimensional surface with $x=\frac{1}{6}$, while right Fig. is the
three-dimensional phase-space portrait. Blue points are the stationary points
at the finite regime. Gree and purple vectors represent the trajectoris with
$\Sigma_{+}=0,~\Sigma_{-}=0$. }%
\label{ff2}%
\end{figure}

In Fig. \ref{ff2} we present the phase-space portrait for the dynamical system
(\ref{ai.15}), (\ref{ai.16}) in the space of variables $\left(  x,\Sigma_{\pm
}\right)  $. It is important to mention that the dynamical system is not
compactified and there exist trajectories which reach infinity. Due to the
nonlinearity and the complexity of the dynamical system, we investigate the
trajectories at infinity numerically.

\begin{figure}[ptbh]
\centering\includegraphics[width=1\textwidth]{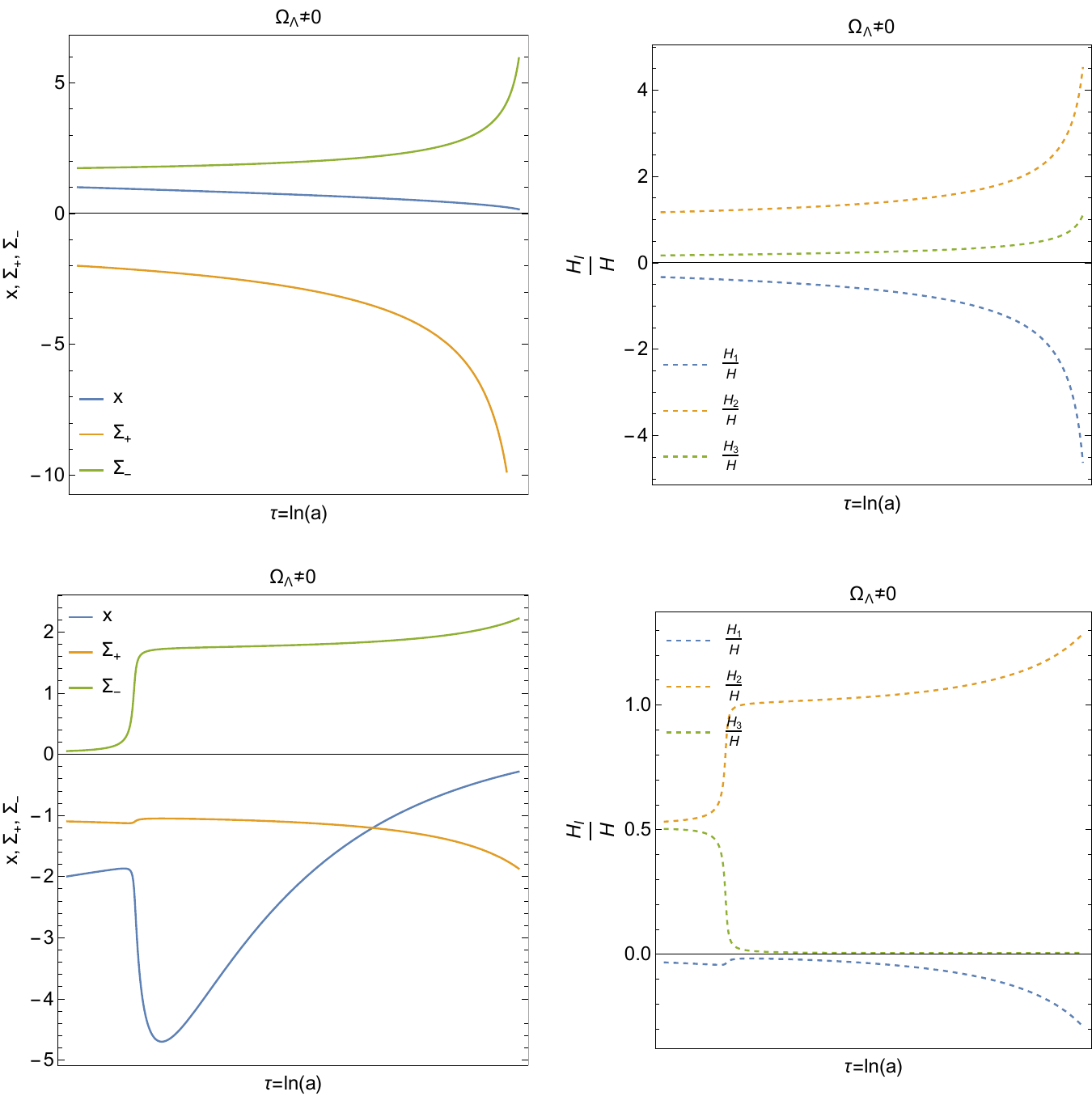}\caption{Qualitative
evolution of the dynamical variables $\left(  x,\Sigma_{\pm}\right)  $ and for
the normalized functions$~\left(  \frac{H_{1}}{H},\frac{H_{2}}{H},\frac{H_{3}%
}{H}\right)  $ for various sets of initial conditions such that the dynamical
system reaches infinity. }%
\label{ff3}%
\end{figure}

Numerical simulations for the dynamical variables $\left(  x,\Sigma_{\pm
}\right)  $ and for the normalized functions$~\left(  \frac{H_{1}}{H}%
,\frac{H_{2}}{H},\frac{H_{3}}{H}\right)  $ are presented in Fig. \ref{ff3},
for initial conditions such that the dynamical system reaches infinity. We can see 
that even if the metric tends to be an isotropic one (shear values tend to zero),
the scalar field does not stabilize, and this regime does not tend to isotropic GR 
de Sitter solution despite the presence of cosmological constant in the action.

\section{5D Spacetime}

\label{sec4d}

For a higher-dimensional geometry the Gauss-Bonnet term contributes additional
dynamical components in the gravitational field equations. Hence, in order to
understand the effects of the dimension on the evolution we consider the same
model within a five-dimensional geometry.

Consider now the five-dimensional anisotropic spacetime
\begin{equation}
ds^{2}=-N^{2}\left(  t\right)  dt^{2}+S_{1}^{2}\left(  t\right)  dx^{2}%
+S_{2}^{2}\left(  t\right)  dy^{2}+S_{3}^{2}\left(  t\right)  dz^{2}+S_{4}%
^{2}\left(  t\right)  dw^{2} \label{bb}%
\end{equation}
where in Misner variables the four scale factors $S_{I}$ are expressed as
\begin{align*}
S_{1}^{2}\left(  t\right)   &  =\exp\left(  2a\left(  t\right)  +3\beta
_{1}\left(  t\right)  \right)  ,\\
S_{2}^{2}\left(  t\right)   &  =\exp\left(  a\left(  t\right)  -\beta
_{1}\left(  t\right)  +2\sqrt{2}\beta_{2}\right)  ,\\
S_{3}^{2}\left(  t\right)   &  =\exp\left(  a\left(  t\right)  -\beta
_{1}\left(  t\right)  -\sqrt{2}\beta_{2}+\sqrt{6}\beta_{3}\right)  ,\\
S_{4}^{2}\left(  t\right)   &  =\exp\left(  a\left(  t\right)  -\beta
_{1}\left(  t\right)  -\sqrt{2}\beta_{2}-\sqrt{6}\beta_{3}\right)  ,
\end{align*}
such that the volume to be $V\left(  t\right)  =N\left(  t\right)  \exp\left(
4a\left(  t\right)  \right)  $ and Hubble function $H=\frac{\dot{a}}{N}$

For assume the gravitational Action Integral (\ref{ai.01}) and we employ the
same procedure as before, where now we introduce the dimensionless variables%
\[
x=\dot{\phi}H,~y=\phi H^{2},~\sigma_{1}=\frac{\dot{\beta}_{1}}{H},~\sigma
_{2}=\frac{\dot{\beta}_{1}}{H},~\sigma_{3}=\frac{\dot{\beta}_{1}}{H}%
,~\Omega_{\Lambda}=\frac{\Lambda}{3H^{2}},~~\tau=\ln a
\]

The field equations are expressed as the following system of
algebraic-differential equations
\begin{equation}
\frac{d\mathbf{\alpha}}{d\tau}=\mathbf{\Psi}\left(  \mathbf{\alpha}\right)  ,
\label{bb.00}%
\end{equation}
where $\mathbf{\alpha}=\left(  x,\psi,\sigma_{1},\sigma_{2},\sigma_{3}\right)
$ and algebraic equation%

\begin{align}
0  &  =8\left(  1-\Omega_{\Lambda}\right)  -\sigma_{1}^{2}-\sigma_{2}%
^{2}-\sigma_{3}^{2}\nonumber\\
&  +2y\left(  2+3\sigma_{1}\right)  \left(  \left(  \sigma_{1}-2\right)
^{2}\left(  \sigma_{1}-6\left(  \sigma_{2}^{2}+\sigma_{3}^{2}\right)
-2\right)  -4\sqrt{2}\sigma_{2}\left(  \sigma_{2}^{2}-3\sigma_{3}^{2}\right)
\right) \nonumber\\
&  -16x\left(  8+2\left(  \sigma_{1}-3\right)  \sigma_{1}^{2}+\sigma_{2}%
^{2}\left(  \sqrt{2}\sigma_{2}-6\right)  -3\left(  2+\sqrt{2}\sigma
_{2}\right)  \sigma_{3}^{2}-3\sigma_{1}\left(  \sigma_{2}^{2}+\sigma_{3}%
^{2}\right)  \right)  . \label{bb.01}%
\end{align}

We compare the latter constraint with that of the four-dimensional manifold
(\ref{ai.16}). We remark that due to the non-boundary term of the Gauss-Bonnet
scalar, in the constraint equation there exist a new variable $y$, with
coefficient $\left(  2+3\sigma_{1}\right)  \left(  \left(  \sigma
_{1}-2\right)  ^{2}\left(  \sigma_{1}-6\left(  \sigma_{2}^{2}+\sigma_{3}%
^{2}\right)  -2\right)  -4\sqrt{2}\sigma_{2}\left(  \sigma_{2}^{2}-3\sigma
_{3}^{2}\right)  \right)  $. However, when $\sigma_{3}=\frac{2-\sigma
_{1}-\sigma_{2}}{\sqrt{6}}$, then the constraint (\ref{bb.01}) is of the form
of (\ref{ai.16}) after a rescale for the anisotropic parameters, and the
previous results are recovered, but such solutions are on an singular surface
and they are always unstable.

\begin{figure}[ptbh]
\centering\includegraphics[width=1\textwidth]{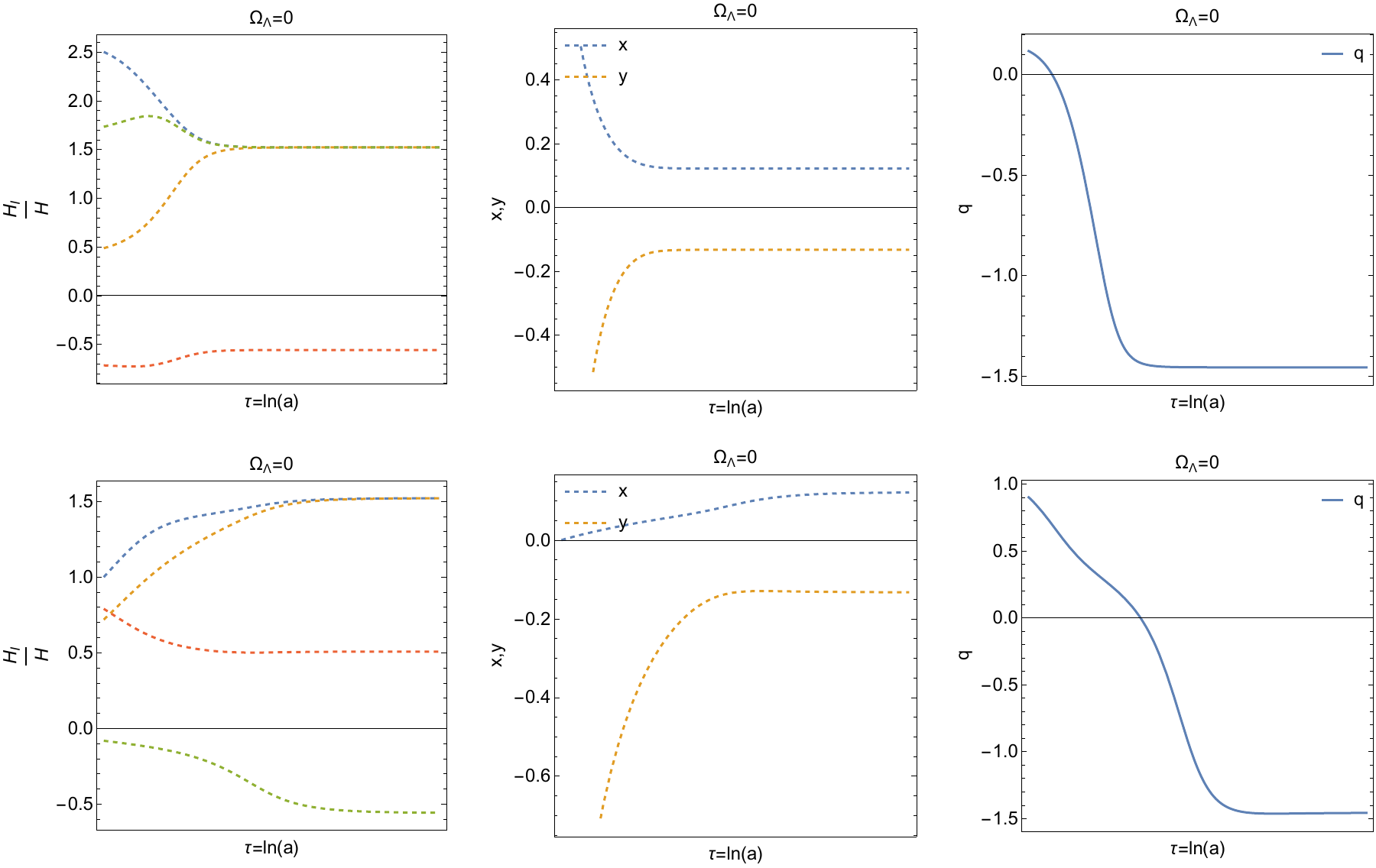}\caption{5D Spacetime:
Qualitative evolution for the four Hubble functions$~\left(  \frac{H_{1}}%
{H},\frac{H_{2}}{H},\frac{H_{3}}{H},\frac{H_{4}}{H}\right)  $, for the scalar
field parameter $\left(  x,y\right)  $ and the deceleration parameter $q$, as
they are given by the solution of the dynamical system (\ref{bb.00}) and
(\ref{bb.01}) for different set of initial conditions and $\Omega_{\Lambda}%
=0$. We observe that the dynamical behaviour of the physical space is
different from the four-dimensional spacetime}%
\label{fig6}%
\end{figure}

\begin{figure}[ptbh]
\centering\includegraphics[width=1\textwidth]{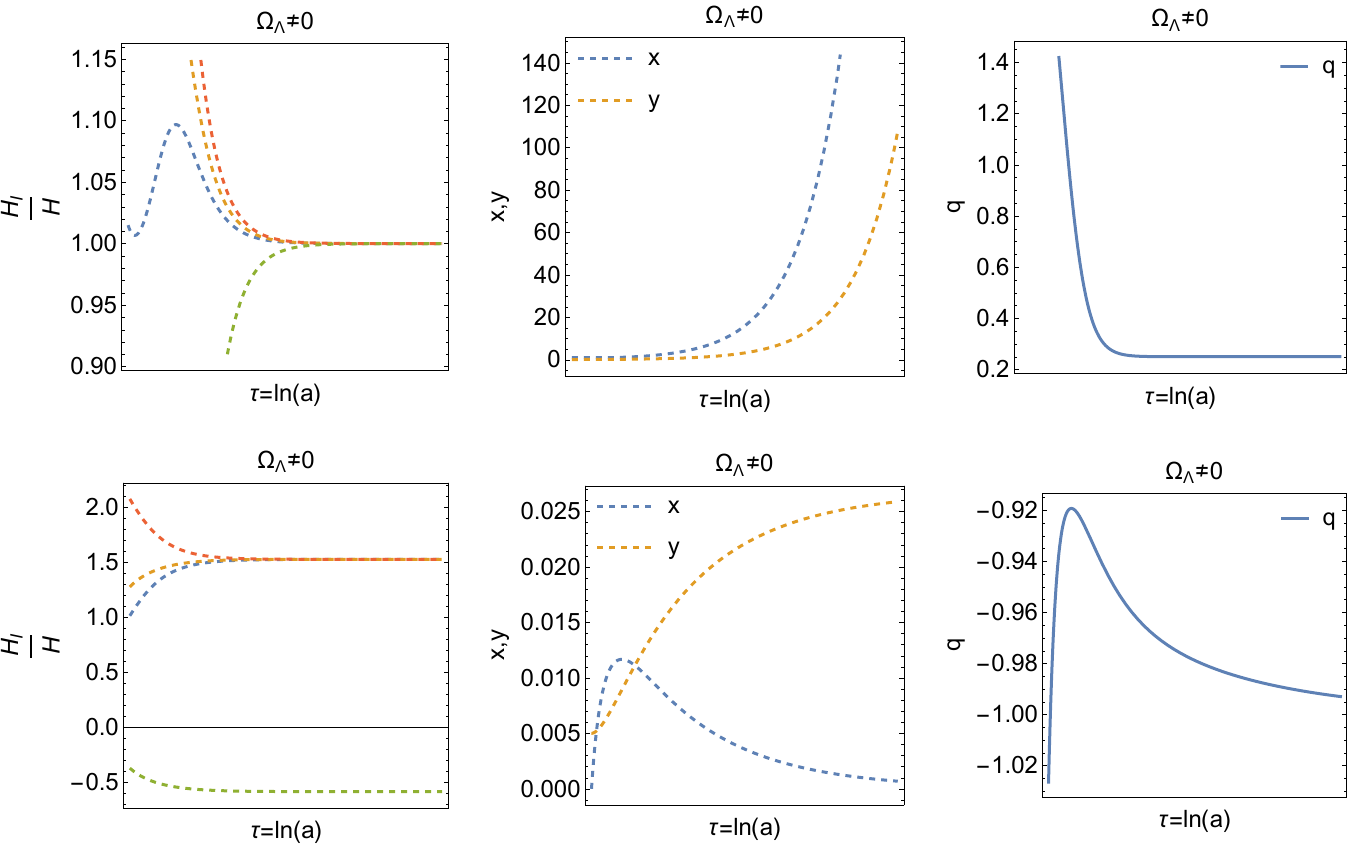}\caption{5D Spacetime:
Qualitative evolution for the four Hubble functions$~\left(  \frac{H_{1}}%
{H},\frac{H_{2}}{H},\frac{H_{3}}{H},\frac{H_{4}}{H}\right)  $, for the scalar
field parameter $\left(  x,y\right)  $ and the deceleration parameter $q$, as
they are given by the solution of the dynamical system (\ref{bb.00}) and
(\ref{bb.01}) for different set of initial conditions and $\Omega_{\Lambda
}\neq0$. We observe that the dynamical behaviour of the physical space is
different from the four-dimensional spacetime}%
\label{fig7}%
\end{figure}

We solve the field equations (\ref{bb.00}), (\ref{bb.01}) numerically and in
Figs. \ref{fig6} and \ref{fig7} we present the qualitative evolution for the
four Hubble function $H_{I}=\frac{S_{I}}{S}$, of the scalar field $x,~y$ and
of the deceleration parameter $q\,$\ for $\Omega_{\Lambda}=0\,,$ and
$\Omega_{\Lambda}\neq0$ respectively.

A trajectory can experience a soft singularity where $\dot H$ diverges while $H$ remaines
finite. This situation leads to diverhence of $q$. In what follows we describe only
non-singular solutions. It is important that we have not got any non-singular solution with
four particular Hubble parameters to be different asymptotically. This means that splitting into
isotropic subspace is still a natural outcome of the cosmological evolution as in the case without 
coupling.

 Apart from splitting, when the cosmological
constant term is introduced, there the exist the limit of isotropization.
 This isotropic solution
is not a de Sitter one since the asymptotic value of $q$ is not equal to $-1$. We see 
also that the value of the scalar field does not stabilise. In this sense this situation is similar 
to 4D case discribed above.

As for solutions with the splitting, we can see 3+1 spatial splitting, and in this solution
$H$ tends to a constant. Note, that though such splitting exists in a pure Gauss-Bonner theory,
it does not satisfy the stability criterion of \cite{Ivashchuk,cir} and have not been found numerically.

Zero $\Lambda$ enables the 3+1 splitting with a running scalar field as well as a new solution with 2+1+1 splitting.
Both solutions have running $\phi$ and non-constant $H$ in the future asymptotic.
 

\begin{figure}[ptbh]
\centering\includegraphics[width=1\textwidth]{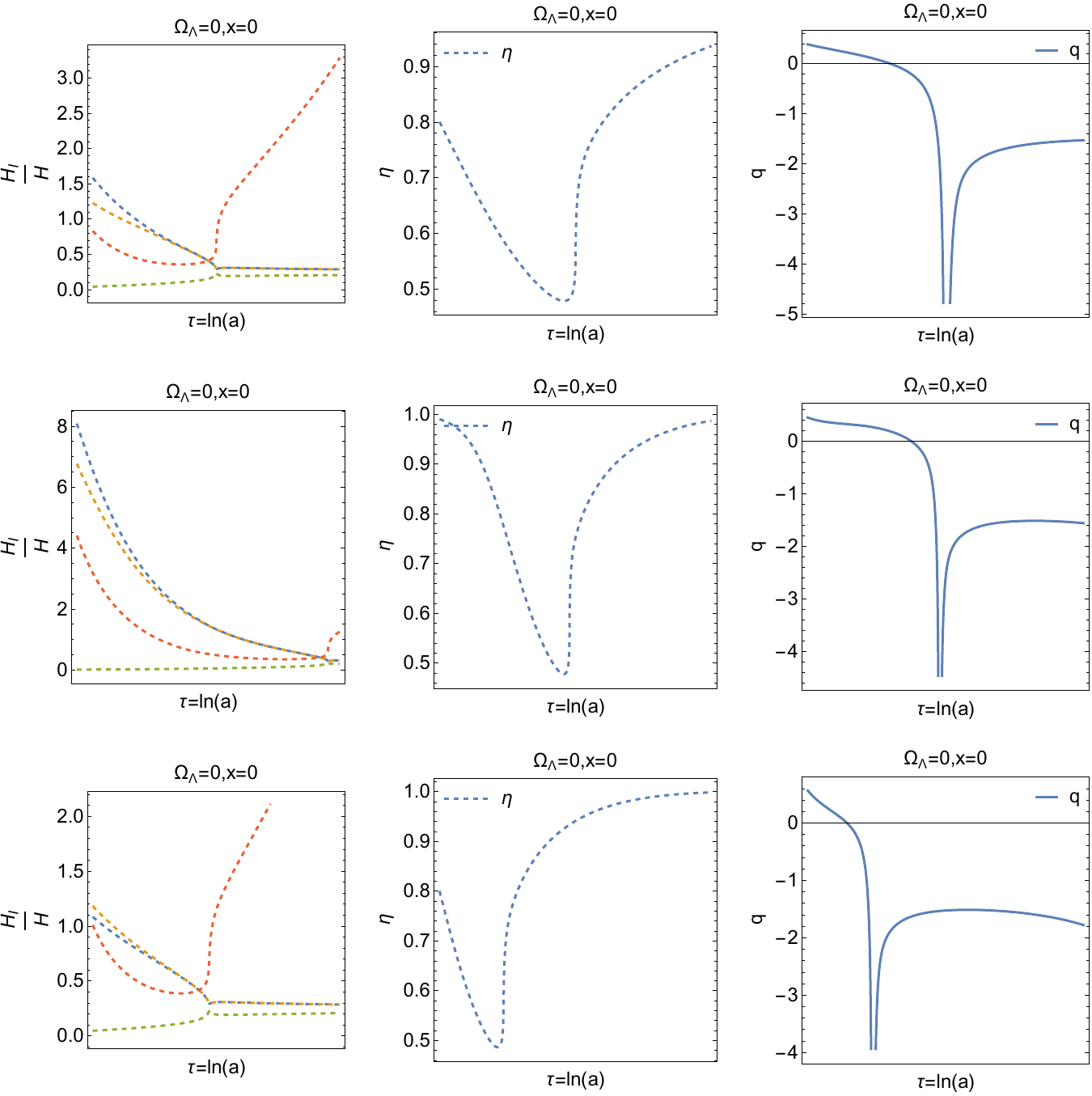}\caption{5D Spacetime:
Qualitative evolution for the pure Gauss-Bonnet gravity, i.e. $\dot{\phi}=0$.
In figues we present the four Hubble functions$~\left(  \frac{H_{1}}{H}%
,\frac{H_{2}}{H},\frac{H_{3}}{H},\frac{H_{4}}{H}\right)  $, parameter $y$, and
the deceleration parameter $q$, as they are given by the solution of the
dynamical system (\ref{bb.00}) and (\ref{bb.01}) for different set of initial
conditions and $\Omega_{\Lambda}=0$. }%
\label{fig8}%
\end{figure}

In order to compare the numerical results with the pure five-dimensional
Gauss-Bonnet gravity, in Fig. \ref{fig8}  we present the
qualitative evolution for the physical parameters on the surface with
$\dot{\phi}=0$, i.e. $x=0$, for $\Lambda\neq0$%.
\begin{figure}[ptbh]
\centering\includegraphics[width=1\textwidth]{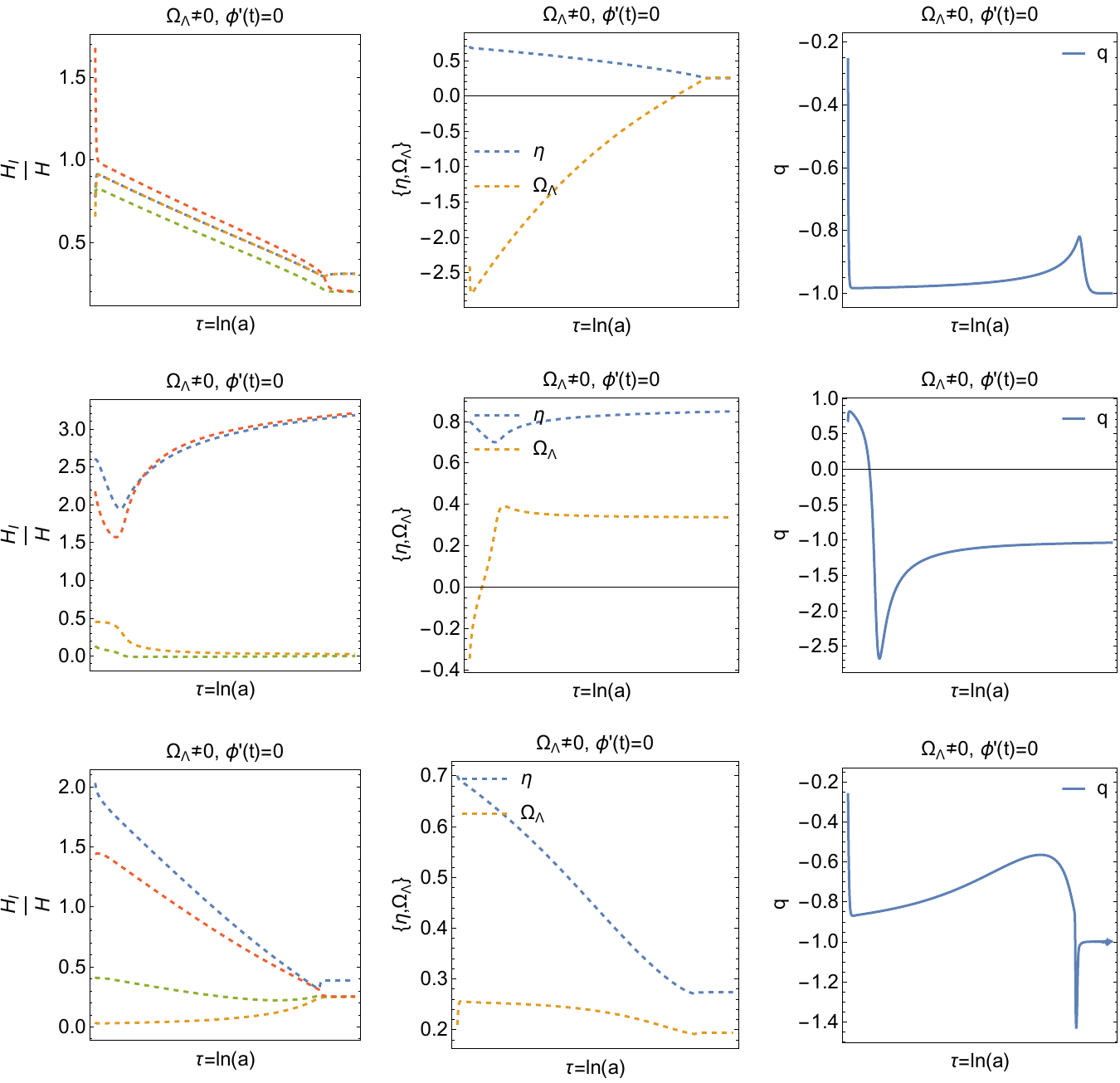}\caption{5D Spacetime:
Qualitative evolution for the pure Gauss-Bonnet gravity, i.e. $\dot{\phi}=0$.
In figues we present the four Hubble functions$~\left(  \frac{H_{1}}{H}%
,\frac{H_{2}}{H},\frac{H_{3}}{H},\frac{H_{4}}{H}\right)  $, parameter $y$, and
the deceleration parameter $q$, as they are given by the solution of the
dynamical system (\ref{bb.00}) and (\ref{bb.01}) for different set of initial
conditions and $\Omega_{\Lambda}\neq0$. }%
\label{fig9}%
\end{figure}
It is known that the only stable splitting in this case is 2+2 splitting,
in our sumulation it was found and presented in Fig.8. No other type of non-singular
trajectories have been detected, is it was expected.

%Indeed, in the absence of the scalar field, the evolution of the
%anisotropies has been investigated in \cite{top2,top1}. The numerical
%simulations in \cite{top2} for the five-dimensional anisotropic geometry
%(\ref{bb}) for the pure Gauss-Bonnet theory lead to the determination of two
%families of trajectories where they have as starting point two families of
%solutions. The standard singularity behaviour which leads to the Kasner
%solution, and a nonstandard type singularity which characterized by the
%divergence of time derivative of the Hubble parameters for its finite value.
%For the first family of trajectories as the universe evolves the Gauss-Bonnet
%term decay and the spacetime reach the five-dimensional Kanser geometry of
%General Relativity.\ These results have been discussed analytically in
%\cite{top1}, where also the differences with the sixth-dimensional anisotropic
%geometry are discussed.

We remark that in the presence of the scalar field the evolution of the
trajectories as they are given in Fig. \ref{fig6}, the asymptotic solution in
the second column reach the limit $x=0$ which means that the nonstandard type
singular solution of \cite{top2} is approached.

\section{Conclusions}

\label{sec5}

In the present paper we have considered possible influence of a scalar field
coupling to Gauss-Bonnet term on the resulting cosmological dynamics. It is
known that in pure vacuum case multidimensional flat Universe for a
considerable part of initial conditions evolves into a product of two isotopic
sub spaces. This property is important for compactification scenarios. So
that, it was reasonable to check if this property still exists for more
complicated theories. It is worth to note that non-minimal coupling with the
Gauss-Bonnet term makes this term to be dynamically important even in 4D
dimensions case, Gauss-Bonnet term itself contributes only starting from 5D
dimensions, so that we consider this case as well. In the present paper we
study only the simplest case when the scalar field coupled to Gauss-Bonnet
term and the kinetic term is absent. The dynamical terms of the scalar field
follows from the integration by parts for the Gauss-Bonnet scalar. Our results
show that in both low-dimension cases considered in the present paper, namely,
4D and 5D dimensions, the nonsingular attractors for cosmological
evolution have similar feathers: in all numerical examples at least two Hubble
parameters tend to be the same.

In particular, for 4D theory the regular outcome is splitting of  spatial part of the metric
into a sum of
isotopic 2-dimensional space and 1-dimensional space. The other possibility is
the isotropic universe. However, in 4D dimensions an isotropic Universe
represents GR asymptotic regime (which corresponds to a constant $\phi$, since
in this case the Gauss-Bonnet term does not contribute to dynamical equations
for a 4D dimensional world) only without $\Lambda$ term, otherwise the fixed
isotropic point does not exists. All our results in 4D dimensions have been
confirmed by analitical studies of the equations of motions.

In 5D theory our results are purely numerical. We have considered a wide
set of initial conditions, and in all cases when future behavior is not
singular we have detected a splitting into isotropic subspaces. If $\Lambda$
is nonzero, the most typical regular outcome is separation of a 3-dim
isotropic subspace. Some initial conditions lead to full isotropisation. Zero
$\Lambda$ adds a possibility of separation of a 2-dim isotropic subspace. No
regular trajectories with four different asymptopic values of Hubble
parameters have been detected.

This means that though final state of the Universe is still a sum of an
isotopic sub space and something else, the set of allowed combinations is
wider that in the pure Einstein Gauss-Bonnet theory, where for four spatial dimensions we have seen only 2
+ 2 spatial dimensions splitting and isotopic nonsingular attractors. Our
results indicate that the 3-dim subspace  separation scenario in the theory where
Gauss-Bonnet term is coupled to a scalar field is possible already in four
spatial dimensions in contrast to pure Einstein Gauss-Bonnet gravity where it
requires at least five spatial dimensions with 3 + 2 splitting. From the other
side, the diversity of possible outcomes can in principle affect the scenario
in bigger number of dimensions and this problem needs further study. Another
possible generalizations of the present work is considering a kinetic term for
the scalar field.

\textbf{Data Availability Statements:} Data sharing is not applicable to this
article as no datasets were generated or analyzed during the current study.

\begin{acknowledgments}
AG was supported by Proyecto Fondecyt Regular 1240247. AP thanks the support
of Vicerrector\'{\i}a de Investigaci\'{o}n y Desarrollo Tecnol\'{o}gico
(Vridt) at Universidad Cat\'{o}lica del Norte through N\'{u}cleo de
Investigaci\'{o}n Geometr\'{\i}a Diferencial y Aplicaciones, Resoluci\'{o}n
Vridt No - 096/2022 and Resoluci\'{o}n Vridt No - 098/2022. AP was partially
supported by Proyecto Fondecyt Regular 2024, Folio 1240514, Etapa 2024. 
\end{acknowledgments}


\begin{thebibliography}{99}                                                                                               %


\bibitem {lov1}D. Lovelock, J. Math. Phys. 12, 498 (1971)

\bibitem {ost1}M. Crisostomi, R. Klein, and D. Roest, J. High Energy Phys. 06,
124 (2017)

\bibitem {pan1}T. Padmanabhan and D. Kothawala, Lanczos-Lovelock models of
gravity, Phys. Reports 531, 115 (2013)

\bibitem {tr1}G. Dotti, J. Oliva and R. Troncoso, Phys. Rev. D 76, 064038 (2007)

\bibitem {ch1}C. Charmousis, J.-F. Dufaux, Class. Quantum Grav. 19, 4671 (2002)

\bibitem {bg1}R.\ Zegers, J. Math. Phys. 46, 072502 (2005)

\bibitem {di1}K.F. Dialektopoulos, D. Malafarina and N. Dadhich, Phys. Rev. D
108, 044080 (2023)

\bibitem {di2}S.G. Ghosh and D.W. Deshkar, Phys. Rev.\ D 77, 047504 (2008)

\bibitem {di3}S. Mukherjee and N. Dadhich, Eur. Phys. J. C 81, 458 (2021)

\bibitem {gbs1}S. Tsujikawa and M. Sami, JCAP 0701, 006 (2007)

\bibitem {gbs2}T. Koivisto and D.F. Mota, Phys. Lett. B 644, 104 (2007)

\bibitem {gbs3}M. Sami, A. Toporensky, P. V. Tretjakov and S. Tsujikawa, Phys.
Lett. B 619, 193 (2005)

\bibitem {gbs4}G. Calcagni, B. de Carlos and A. De Felice, Nucl. Phys. B 752,
404 (2006)

\bibitem {gbs5}D. D. Doneva, K.V. Staykov and S. S. Yazadjiev, Phys. Rev. D
99, 104045 (2019)

\bibitem {gbc1}M.A. Garcia-Aspeitia and A. Hern\'{a}ndez-Almada, Phys. Dark
Univ. 32, 100799 (2021)

\bibitem {top2}R. Chingangbam, M. Sami, P. V. Tretyakov and A.V. Toporensky,
Phys.\ Lett. B 661, 162 (2008)

\bibitem {top1}S. A. Pavluchenko and A. V. Toporensky, Mod. Phys. Lett. A 24,
513 (2009)

\bibitem {dd1}D. Glavan and C. Lin, Einstein-Gauss-Bonnet gravity in
4-dimensional space-time, Phys. Rev. Lett. 124, 081301 (2020)

\bibitem {dd2}P.G.S\ Fernandes, P. Carrilho, T. Clifton and D.J. Mulryne, The
4D Einstein-Gauss-Bonnet Theory of Gravity: A Review, Class. Quantum Grav. 39,
063001 (2022)

\bibitem {gbm01}B. Li, J.D. Barrow and D.F.\ Mota, Cosmology of modified
Gauss-Bonnet gravity, Phys. Rev. D 76, 044027 (2007)

\bibitem {gbm03}S. Nojiri, S.D. Odintsov, V.K.\ Oikonomou and A.V. Popov,
Ghost-free F(R,G) gravity, Nuclear Phys. B 973, 115617 (2021)

\bibitem {agb1}D. Chirkov, A. Giacomini, A. Toporensky and P. Tretyakov, Gen.
Rel. Grav. 56, 110 (2024)

\bibitem {gg1}M. Motaharfar and H. R. Sepangi, Eur. Phys. J. C 76, 646 (2016)

\bibitem {gg2}P. Kanti, R. Gannouji and N. Dadhich, Phys. Rev. D 92, 041302 (2015)

\bibitem {gg2a}I. V. Fomin, Phys. Part. Nucl. 49, 525 (2018)

\bibitem {gg3}R.A. Konoplya, T. Pappas and A. Zhidenko, Phys. Rev. D 101,
044054 (2020)

\bibitem {gg4}N. Chatzarakis and V.K. Oikonomou, Annals of Physics 419, 168216 (2020)

\bibitem {gg5}S.D. Odintsov, V.K. Oikonomou and F.P. Fronimos, Annals of
Physics 420, 168250 (2020)

\bibitem {gg6}H.M. Sadjadi, Phys. Lett. B. 850, 138508 (2024)

\bibitem {gg7}G. Hikmawan, J. Soda, A. Suroso and F. P. Zen, Phys. Rev. D 93,
068301 (2016)

\bibitem {gg8}B. Micolta-Riascos, A.D. Milano, G. Leon, B. Droguett and E.
Gonzalez, Fractal Fract. 8, 626 (2024)

\bibitem {gg9}A.D Milano, C. Michea,\ G. Leon and A. Paliathanasis, Phys. Dark
Univ. 46, 101589 (2024)

\bibitem {gg10}A. Paliathanasis, Gen.\ Rel. Grav. 56, 84 (2024)

\bibitem {gg11}R. D' Agostino, O. Luongo and S. Mancini, Eur. Phys. J. C 84, 1060\ (2024)

\bibitem {gg12}M.A.S. Pinto and J.L. Rosa, arXiv:2411.04066 (2024)

\bibitem {g11}D. Chirkov, S. A. Pavluchenko and A. Toporensky, Gen.\ Rel.
Grav. 46, 1799 (2014) 

\bibitem {g12}E.O. Pozdeeva, M.\ Sami,\ A.V. Toporensky and S. Yu Vernov,
Phys. Rev. D 100, 083527 (2019)

\bibitem {g13}E.O. Pozdeeva, M.R. Gangopadhyay, M Sami, A.V. Toporensky and S.
Yu Vernov, Phys.\ Rev. D 043525 (2020)

\bibitem {g14}E.O. Pozdeeva, M.A. Skugoreva, A.V. Toporensky and S. Yu Vernov,
JCAP 09, 050 (2024)

\bibitem {cop1}E.J. Copeland, A.R. Liddle and D. Wands, Phys. Rev. D 57, 4686 (1998)

\bibitem {Ivashchuk} V.D. Ivashchuk, Eur. Phys. J. C 76 431 (2016)

\bibitem {cir} D. M. Chirkov, A. V. Toporensky, Gravitation and Cosmology 23, 4, 359 (2017)
\end{thebibliography}
\end{document}